\RequirePackage{fix-cm}

\documentclass[smallextended]{svjour3}       
\smartqed  
\usepackage{graphicx}
\usepackage{subfigure}

\journalname{XXX}
\begin{document}

\title{ Controlled Alternate Quantum Walk based Block Hash Function
}


\author{Dan Li         \and
       Panpan Ding    \and
       Yuqian Zhou     \and
       Yuguang Yang
}


\institute{Dan Li        \and
       Panpan Ding     \and
       Yuqian Zhou          \at
              College of Computer Science and Technology, Nanjing University of Aeronautics and Astronautics, Nanjing, China \\
              \email{lidansusu007@163.com}          
           \and
       Yuguang Yang  \at
              Faculty of Information Technology, Beijing University of Technology, Beijing, China
}

\date{Received: date / Accepted: date}

\maketitle

\begin{abstract}
The hash function is an important branch of cryptology. Controlled quantum walk based hash function is a kind of novel hash function, which is safe, flexible, high-efficient, and compatible. All existing controlled quantum walk based hash functions are controlled by one bit message in each step. To process message in batch amounts, in this paper, controlled alternate quantum walk based block hash function is presented by using the time-position-dependent controlled quantum walks on complete graphs with self-loops. The presented hash function accelerate the hash processing dramatically, so it is more high-efficient. 

\end{abstract}

\section{Introduction}
\label{sec1}

The hash function is an important branch of cryptology. The Hash function includes all irreversible functions that can be used  to map arbitrary size data onto  fixed-size data. The values returned by a hash function are called  hash values. A hash function allows one to easily verify whether some input data map onto a given hash value, but if the input data is unknown it is deliberately difficult to reconstruct it by knowing the stored hash value. There are many theoretical studies about classical hash functions and mature hash functions such as MD5, SHA1, and SHA512. Furthermore, hash functions can be used in message authentication code and public key infrastructure. These hash functions are generally constructed based on mathematics complexity and thus they are computationally secure.

The rapid development of quantum computation brings a great challenge to cryptology. Shor's integer factorization algorithm collapses the security of many public-key cryptography protocols. Grover search algorithm threatens the security of all symmetric cryptography protocols. Therefore, how to develop secure cryptology is very important. Except for post-quantum cryptography, quantum cryptography \cite{QC} is the principal method, for instance, quantum key distribution. Quantum computation plays an unignorable  act in the field of quantum cryptography.

Quantum walk, the quantum counterparts of classical random walks, is an underlying mathematical model in realizing quantum computation. Alternate quantum walks, quantum walks with memory, quantum walks on different kinds of graphs are presented in Ref. \cite{QWM,QWMS,QWMP,QWCayley,QWHypergraph}  for different purposes. Quantum walk has wide applications in quantum computation and quantum communication, such as database searching, element distinctness, graph isomorphism testing, and quantum network communication.

In 2013, Li et. al. \cite{LiCIQW,LiCIQW2} presented the two-particle controlled interacting quantum walk (CIQW) and CIQW-based quantum hash function. A comprehensive analysis of CIQW proved that it is suitable for designing quantum hash functions. This quantum hash function guarantees the security of the hash function by the irreversibility of measurement rather than hard mathematic problems. This work opens the door to the research of the controlled quantum walk (CQW)-based hash function.

In the next years, the CQW-based hash function attracts much attention from researchers. Many kinds of controlled quantum walk on diffident graphs are introduced for building different CQW-based hash functions. Yang et. al. \cite{YangApp} improved the CIQW-based quantum hash function and found its applications in the privacy amplification process of quantum key distribution, pseudo-random number generation, and image encryption. After that, Yang et.al \cite{YangJohnson} present a quantum hash function based on the controlled quantum walk on Johnson graphs, which has a lower collision rate and quantum resource cost. The CQW-based hash function with a controlled quantum walk on cycles is also presented by Yang et.al \cite{YangSimple} in 2018. The CQW-based hash function with a controlled quantum walk on cycles with two coins is presented by Yang et.al \cite{YangTwo} in 2019. Besides, the CQW-based hash function with a decoherent quantum walk on a cycle is shown by Yang et.al \cite{YangDecoherence} in 2021.

In 2018, Li et.al. \cite{LiCAQW} presented the controlled alternate quantum walk (CAQW)-based quantum hash function. The controlled alternate quantum walk saves quantum resource costs dramatically. Then, the controlled alternate quantum walk is used in constructing pseudo-random number generator and quantum color image encryption \cite{AAAImage,LiImage1,LiImage2,HaoImage}. Zhou presented the hash function based on controlled alternate quantum walks with memory in 2021 \cite{ZhouMemory}. They claim that the proposed hash function has near-ideal statistical performance.

Although the research on CQW-based hash function thriving, how to improve the efficiency is an important research orientation. In each step of the existing CQW till now, the coin operator is controlled by a one-bit message. That is not efficient enough especially when the controlled quantum walk is executed on a quantum computer, due to the difficulties in implementing a quantum computer.

To speed up the enforcement efficiency of the CQW-based hash function further, in this paper, the time-position-dependent  CAQW on complete graphs with self-loops is introduced to build the controlled alternate quantum walk-based block hash function (CAQWBH). CAQWBH  could accelerate the hash processing dramatically by processing a message in batch amounts. Complete graphs with self-loops are chosen for constructing hash functions that satisfy the avalanche characteristic.

The paper is structured as follows. In Sect.\ref{sec2} and Sect.\ref{sec3}, a time-position-dependent  CAQW on complete graphs with self-loops and CAQWBH are presented respectively.  The security analysis and statistical performance	of CAQWBH are discussed in Sect.\ref{sec4} and Sect. \ref{sec5}. In Sect. \ref{sec6}, extensions of CAQWBH, including  message verification code and pseudo-random number generator are presented.
Finally, a discussion is given  in Sect.\ref{sec7}.

\section{A Time-position-dependent Controlled Alternate Quantum walk on complete graphs with self-loops}
\label{sec2}

In this section, a  time-position-dependent CAQW on a complete graph with self-loops is introduced for the construction of CAQWBH. Complete graphs with self-loops are chosen for the reason that  CAQWBH must satisfy the avalanche characteristic.

Suppose $G_N$ is the complete graph with $N$ vertexes and $N$ self-loops. Here $N=2^q$. The time-position-dependent CAQW on $G_N$ takes place in the product space $H_p\otimes H_c$. $H_p$ is spanned by $\{|x\rangle, x\in \{0,\cdots, N-1\}\}$. Furthermore, $H_p$ can also be described as  $\{|x_q,\cdots,x_1\rangle, x_i\in \{0, 1\} $. $H_c$ is spanned by $\{|c\rangle, c\in \{0,1\}\}$.

Let $|x,c\rangle$ or $|x_q,\cdots,x_1,c\rangle$  be a basis state, where $x$ and $c$ represent the position and the coin state of the walker respectively. In each step of the time-position-dependent CAQW on  $G_N$, the evolution of the whole system can be described by the global unitary operator, denoted by $U$,

\begin{eqnarray}\label{E201}
  U(t)=S_5CS_4CS_3CS_2CS_1C.
\end{eqnarray}

The coin operator $C=(\sum_x |x\rangle\langle x|\otimes C_2(x,t))$ is a time-position-dependent unitary operator. The 2-dimensional operator $C_2(x,t)$ is customarily described as follows.
\begin{eqnarray}\label{E202}
  C_{2,i}=
\left(
  \begin{array}{cccc}
    cos(\theta_i) & sin(\theta_i)  \\
    sin(\theta_i) & -cos(\theta_i)  \\
  \end{array}
\right)
\end{eqnarray}
Two parameters $\theta_0, \theta_1$ are selected from $(0, \pi/4) \cup (\pi/4, \pi/2)$ to construct two unitary operators $C_{2,0}$ and $C_{2,1}$ respectively. For each $t$, $C(t)$ is controlled by a $N$-bit binary string, i.e. message.

The shift operators $S_i$ is defined by
\begin{eqnarray}\label{E203}
  S_i=\sum_{x,c}   |\ x_i+c (mod\ 2)\ \rangle\langle\  x_i|\otimes |c\ \rangle\langle\ c|.
\end{eqnarray}  As a result, $U(t)$ is controlled by a $N$-bit message for each $t$. If the length of the message is not a multiple of  $N$, the Hadmard matrix is the default choice for $C_{2,i}$.

Then, the final state can be  expressed by
\begin{eqnarray}\label{E204}
  |\psi_{t}\rangle=U(t)\times \cdots \times U(1)|\psi_{0}\rangle,
\end{eqnarray}where $|\psi_{0}\rangle$ is the initial state of the total quantum system. Hence the probability of finding the walker at position $x$ after $t$ steps is
\begin{eqnarray}\label{E205}
  P(x,t)=\sum_{c}|\ \langle x,c|U(t)\times \cdots \times U(1)|\psi_{0}\rangle\ |^2.
\end{eqnarray}
Due to the evolution of quantum walks is obtained by taking the Kronecker product of sparse matrices, being executed on a quantum computer or being simulated on a classical computer are efficient and undemanding.

It needs to be emphasized that if the underlying graph is not a complete graph, two different messages may lead to the same probability distribution on purpose.

\section{The Controlled Alternate Quantum Walk-Based  Block Hash Function}
\label{sec3}

CAQWBH  is constructed by running the time-position-dependent CAQW on $G_N$. The process of CAQWBH is described as follows:

\begin{enumerate}{}{}
\item{Select the parameters $(N, k, (\theta_1,\theta_2), \alpha_i)$. $\theta_1,\theta_2\in(0,\pi/2)$, $i\in \{0,\cdots, N-1\}$ and $\sum_{i}|\alpha_i|^2=1$.  $\theta_1,\theta_2$ are the parameters of the two coin operators respectively.}
\item{Initialize the quantum system. The initial state is $|0\rangle(\sum_i\alpha_i|i\rangle)$. Then run the time-position-dependent CAQW on $G_N$ one step under the control of a binary string which is consisted of $N$ $0$s. The purpose of this step is to make sure the quantum system begins with a superposition state that all probability amplitudes are nonzero.}
\item{Run the  time-position-dependent CAQW on $G_N$ to get the final state under the control of the message. The default option for $C_2$ is the Hamdmard matrix, if the length of the message is not a multiple of $N$.}
\item{Post-processing of the probability distribution to get the hash value.  Multiply all values in the resulting pseudo-probability distribution by $10^k$. Then retain the remainders of modulo $2^k$ to form a binary string as the hash value. The bit length of the hash value is $N\times k$.}
\end{enumerate}

\begin{table}
  \centering
  \caption{The relation between the parameters and the length of the hash value}\label{Table1}
\begin{tabular}{|c|c|c|c|}
\hline
Hash Instances & $N$ & $k$ & message length at time $t$  \\
\hline
CAQWBH-256 &$N=32$ &  $k=8$  &  $32t$  \\
\hline
CAQWBH-512 &$N=64$ &  $k=8$  &  $64t$  \\
\hline
\end{tabular}
\end{table}

As a hash function, apart from satisfying the basic requirements, CAQWBH is flexible, high-efficient, and compatible. Concrete analysis is as follows.

Flexible: CAQWBH with the hash value of different lengths are easy to construct by changing the parameters $N$ and $k$ rather than the structure of CAQWBH. The relation between the parameters and the length of the hash value is shown in Table. \ref{Table1}.

High-efficient: In each step of the time-position-dependent CAQW on $G_N$, the coin operator is controlled by a $N$-bit message. Therefore, CAQWBH can  process message in batch amount, which could accelerate the hash processing dramatically compare to other CQW-based hash functions.

Compatible: CQW-based hash function is compatible to be executed on a quantum computer or a classical computer. So is CAQWBH. To get the accurate probability distribution, the time-position-dependent CAQW needs to be executed many times on a quantum computer. At the same time, CAQWBH is safe and even more high-efficient on a classical computer for the present. Taking the difficulty of the commercialization of the quantum computer into consideration. CAQWBH is practical in current hardware technology.

\section{Security analysis}
\label{sec4}

The security of CAQWBH is based on the irreversibility of measurement and the modulo arithmetic.

\begin{enumerate}{}{}
\item{By using the modulo operator, the probability distribution is transformed to the hash value. This process is irreversible because it is a many-to-one relationship. The probability of transforming the hash value back to the right probability distribution is approximately 0. It is the first shield to protect the message from unauthorized persons.}
\item{The second shield of CAQWBH is the irreversibility of measurement. The final state of the time-position-dependent CAQW is in the following form $|\psi_{t}\rangle=\sum_{x,c}\lambda_{x,c}|\ x,c \ \rangle$. This state is a pure state and is linear with the initial state. The probability distribution is the sum of squares of the probability amplitudes, i.e. $P(x)=\sum_c |\ \lambda_{x,c}\ |^2$.  As a result, the final state is easy to obtain by a quantum computer or a classical computer. However, the measurement processing breaks the linearity between the final state and the initial state. And the chance of decomposing the probability as the sum of squares of probability amplitudes is 0.}
\end{enumerate}

Above two irreversible computational processes guarantee that it is impossible to backtrack the final state from the hash value. Therefore, it is impossible to obtain the message, let alone find another message who can generate the same hash value, even though the initial state is public.


\section{Statistical Performance Analysis}
\label{sec5}

In this section, we performed several hash tests to evaluate the performance of  CAQWWBH. $N=8$, $k=8$, $cos(\theta_1)=3/5$, $cos(\theta_2)=8/17$ are chosen in this part. The initial state is $|0\rangle_p|0\rangle_c$. And the hash value we consider here is 256 bits. The results show that CAQWBH have outstanding statistical performance.

\subsection{Sensitivity of hash value to message}
\label{sec51}

Let $Mes1$ be the original message.  $Mes2$, $Mes3$, $Mes4$ represent the messages with tiny modifications of  Mes1. The results listed below show the high sensitivity to the message and the tiny changes.

Condition 1: Random choose an original message $Mes1$;

Condition 2: Flip a bit of $Mes1$ at a random position and then obtain the
modified message $Mes2$;

Condition 3: Delete a  bit from $Mes1$ at a random position and then obtain the
modified message $Mes3$;

Condition 4: Insert a random bit into $Mes1$ at a random position and then obtain the
modified message $Mes4$;

The sensitivity of the hash value to message is assessed by comparing the hash values of the modified messages with that  of the original one. And the corresponding 256-bit hash values in the hexadecimal format are given by:

Condition 1: B267FDFA71168265F4AF9B71FFDB446F

             51C116F2D74B2DBC3CF53C0A16E2821E;

Condition 2: 199F165B52D947673372A5EB2516F061

             BD3CD821C0D00DC348FEDED083BDC4DF;

Condition 3: 58E54047F19DA2459BE3B39809B04586

             33C189DE0492068665CB6B8D29B28367;

Condition 4: 086B23C50D8D9E2E5931A08F3E0D2D88

             112EE1663C403D7D1E68142BDE7A8B0F.

\begin{figure}[!htb]
 \begin{center}
 \caption{Plots of the 256-bit Hash Value C1, C2, C3 and C4}\label{Fig001}
 \includegraphics[width=10cm]{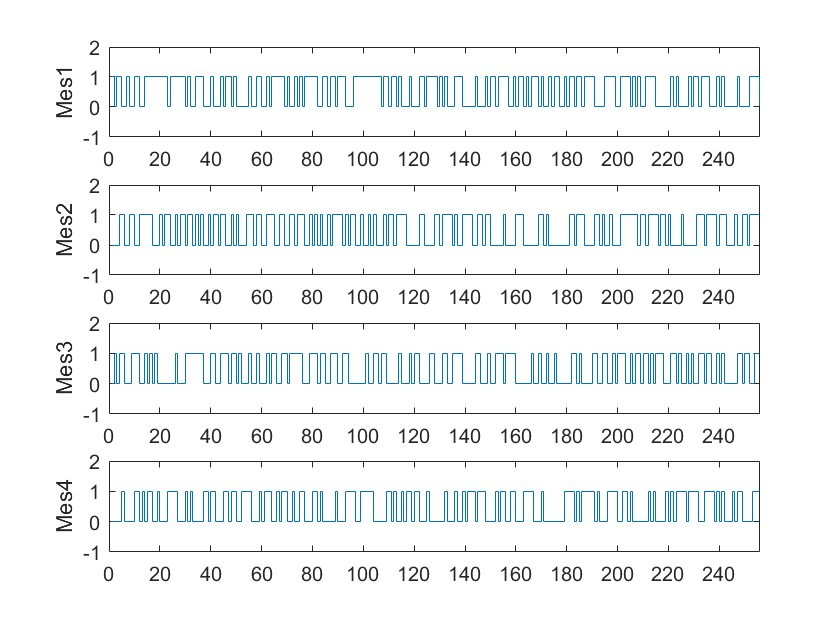}\\
  \end{center}
  \renewcommand{\figurename}{Fig.}
\end{figure}

The plots of the hash values in the binary format are shown  in Fig.\ref{Fig001}. It clearly indicates that any tiny modification to the message will cause a substantial change in the final hash value. A similar result can be obtained using any other instances of CAQWBH. Therefore, the output hash value of CAQWBH is highly sensitive to its input message.

\subsection{Statistical analysis of diffusion and confusion}
\label{sec52}

The diffusion and confusion tests are performed as follows:

(1) Random choose an original message $Mes1$ and generate the corresponding hash value;

(2) Flip a bit of $Mes1$ at a random position to obtain $Mes2$ and generate a new hash value;

(3) Compare the two hash values and count the number of changed bits at the same location called $B_i$;

(4) Repeat steps (1) to (3) $T$ times.

The diffusion and confusion properties of CAQWBH are assessed based on the following indicators:

Minimum changed bit number $B_{min}=min(\{B_i\}^T_1)$;

Maximum changed bit number $B_{max}=max(\{B_i\}^T_1)$;

Mean changed bit number $\overline{B}=\sum^T_{i=1}B_i/T$;

Mean changed probability $P=(\overline{B}/128)\times100\%$;

Standard deviation of the changed bit number $\bigtriangleup B=\sqrt{\frac{1}{T-1}\sum^T_{i=1}(B_i-\overline{B})^2}$;

Standard deviation of the changed probability $\bigtriangleup P=\sqrt{\frac{1}{T-1}\sum^T_{i=1}(B_i/128-P)^2}\times100\%$.

The diffusion and confusion tests are perforemed $T=10000$ times as shown in Table \ref{Table2}. We concluded from the tests that the mean changed bit number $\overline{B}$ and the mean changed probability $P$ are close to the ideal value 64 and $50\%$ respectively. $\bigtriangleup B$ and $\bigtriangleup P$ are very little, $B_{min}$ and $B_{max}$ are around 64, so that it demonstrates the stability of diffusion and confusion. The excellent statistical effect ensures that it is impossible to forge plaintext-cipher pairs given known plaintext-cipher pairs.

\begin{table}
  \centering
  \caption{Static Number of Changed Bit $B$}\label{Table2}
\begin{tabular}{|c|c|c|c|c|c|c|c|c|c|}
\hline
T=10000 & $\overline{B}$ & $P(\%)$ & $\bigtriangleup B$ &  $ \bigtriangleup P$ & $ B_{min} $ & $ B_{max}$ \\
\hline
CAQWBH-256 & 128.11 &  50.05 &  7.90 &  3.09 & 99 &  156  \\
\hline
\end{tabular}
\end{table}

\subsection{Uniform distribution on hash space}
\label{sec53}

\begin{figure}[!ht]
 \begin{center}
 \caption{Uniform Distribution on Hash Space.} \label{Fig002}
  \includegraphics[width=10cm]{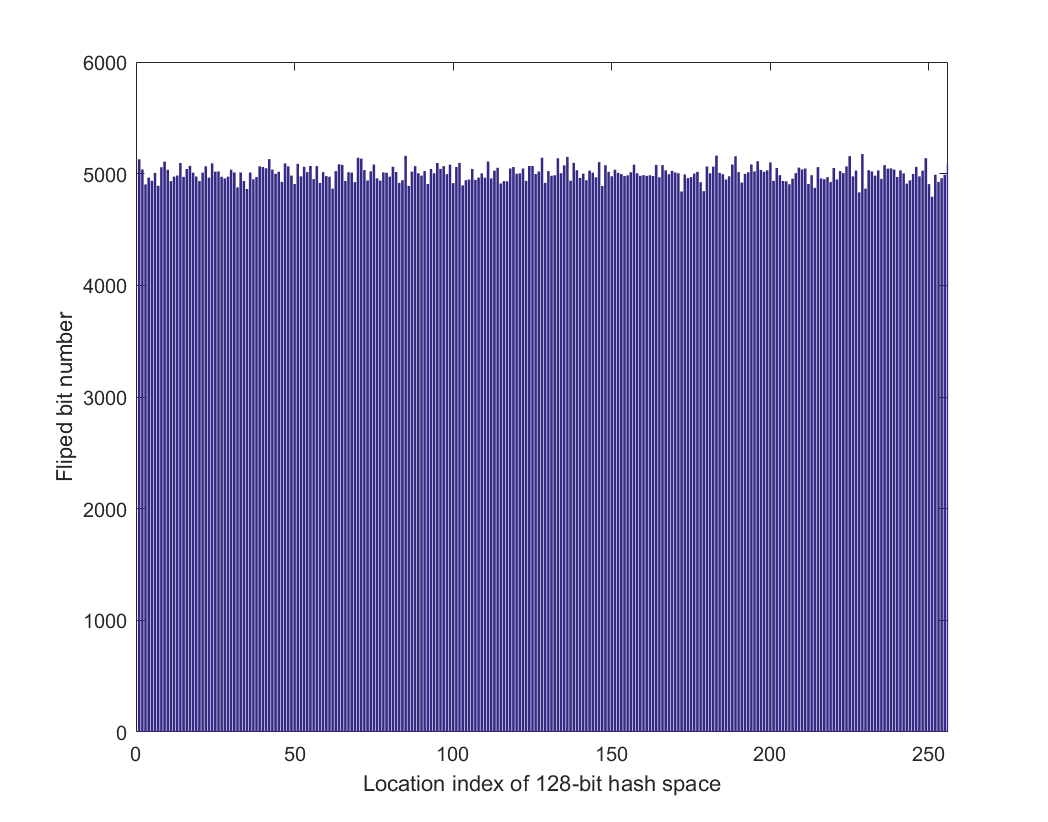}\\
  \end{center}
  \renewcommand{\figurename}{Fig.}
\end{figure}

In order to check the distribution capacity in hash space, we generated two hash values according to the method described in previous subsection and then counted the number of the changed bits at each location. The statistical results for $T = 10,000$ are shown in  Fig.\ref{Fig002}. The mean of the changed bit number 5004.58  is very close to the ideal value 5000, which accounts for half of the test times. It can be concluded that the hash value is distributed uniformly in the hash space as all the changed bit numbers are around the ideal value. Obviously, this demonstrates the resistance against statistical attack.

\subsection{Collision analysis}
\label{sec54}
It is hard to provide a mathematical proof on the capability of collision resistance of chaotic hash functions. Thus, we performed  tests for analyzing collision resistance. Let $W_T(\omega)$ the number of draws  on which the hash values of the original and modified message contain $\omega$ bytes with the same value at the same location. It is a common indicator of the collision resistance property. If the experimental result of $W_T(\omega)$ is very close to the theoretical value, then the hash function could be regarded as having a good property of collision resistance.

The collision tests are performed as follows:

(1) Random choose an original message $Mes1$ and generate the corresponding hash value in byte format;

(2) Flip a bit of $Mes1$ at a random position to obtain $Mes2$ and generate a new hash value in byte format;

(3) Compare the two hash values and count the number of same bytes at the same location called $\omega$;

(4) Repeat steps (1) to (3) $T$ times.

From the meaning of $\omega$, \begin{eqnarray}\label{E501}
 \omega=\sum^{T}_{i=1}\delta(e_i-e'_i)
\end{eqnarray}where $\delta(x)$ is  the Dirac  delta function. $e_i$ and $e'_i$ are the $i$th entries of the original and the new hash value in byte format respectively.

Then, through $T=10000$ independent tests, $W_T(\omega)$ can be computed according to the following formulas:\begin{eqnarray}\label{E502}
 && W_T(\omega)=T\times Prob\{\omega\}=T\frac{n!}{\omega!(n-\omega)!}\left( \frac{1}{2^8}\right)^\omega \left(1- \frac{1}{2^8}\right)^{n-\omega}
\hspace{1mm}
\end{eqnarray}where $\omega=0,1,\cdots, n$. And $n=128/8=16$ in CAQWBH. After running the above tests, the experimental values and the  experimental values of  $W_N(\omega)$ in the proposed hash function are shown in Table \ref{Table3}. The experimental values of $W_N(\omega)$ are similar to the theoretical values.

\begin{table}[!htb]
  \centering
  \caption{ Comparison of Experimental Values and Theoretical Values of $W_N(\omega)$}\label{Table3}
\begin{tabular}{|c|c|c|c|c|c|c|}
\hline
 & $\omega=0$ & $\omega=1$ & $\omega=2$  & $\omega=3$ & $\omega=4$ &$\omega=5\cdots25$\\
\hline
Experimental Values of $W_N(\omega)$ &  8823 & 1101 &  72 & 4 &  0  &  0 \\
\hline
\ \ Theoretical Values of $W_N(\omega)$ &  8822.81 & 1107.18 &  67.30  &  2.64 &  0.08 & 0\\
\hline
\end{tabular}
\end{table}

\subsection{Resistance to birthday attack}
\label{sec55}

Birthday attack implies a lower bound of the length of hash value. The length of the hash value we considered here is $256=32\times8 $ bits. Therefore, it needs  $2^{n/2}=2^{128}\approx 3.4028\times 10^{38}$ trials ($n$ is the size of hash value) to find two messages with identical hash values with a probability of 1/2. Furthermore, CAQWBH-256 can be easily extended to be $512$ bits or more. Therefore, the results of the tests, the size of the hash value, and the collision resistance of the proposed CAQWBH suggest that the birthday attack is almost impossible and that the proposed algorithm is resistant against this type of attack.

In conclusion, CAQWBH has outstanding statistical performance. Compare with other CQW-based hash functions, there is no apparent distinction between them. If the  differences of statistical performance have to be taken into account, the most important factor to the statistical performance is the length of hash value  rather than the varieties of CQW-based hash function.

\section{Extensions of CAQWBH}
\label{sec6}

The structure of CAQWBH could also be used in designing message authentication code(MAC) or pseudo-random number generator(PRNG). The kernel of the corresponding MAC and PRNG is CAQWBH. The security and efficiency  of CAQWBH makes sure that the corresponding CAQWBH-based MAC and CAQWBH-based PRNG are safe enough. Furthermore, the CAQWBH-based MAC and CAQWBH-based PRNG could be used in designing cryptographic protocols.

\subsection{The CAQWBH-based Message Authentication Code}

MAC needs the key to control the  generating process. In order to construct a hash-based MAC, the existing hash-based MAC uses the key to build the new message for hash processing.

Nevertheless, the initial state of CAQW naturally can be considered as the key to the CAQWBH-based MAC. What we need to do is set the initial state as $|0\rangle(\sum_i\alpha_i|i\rangle)$, which is private. Then  CAQWBH becomes  CAQWBH-based MAC. The process of CAQWBH-based MAC is shown as follows.

\begin{enumerate}{}{}
\item{Select the parameters $(N, k, (\theta_1,\theta_2), \alpha_i)$. $\theta_1,\theta_2\in(0,\pi/2)$, $i\in \{0,\cdots, N-1\}$ and $\sum_{i}|\alpha_i|^2=1$.  $\theta_1,\theta_2$ are the parameters of the two coin operators respectively. Among the parameters, $\alpha_i$s can be set as private, which is the $key_1$.}
\item{Initialize the quantum system. The initial state is $|0\rangle(\sum_i\alpha_i|i\rangle)$. Then run the time-position-dependent CAQW on $G_N$  under the control of a binary string, which is the $key_2$.}
\item{Run the  time-position-dependent CAQW on $G_N$ under the control of the message. The default option for $C_2$ is the Hamdmard matrix, if the length of the message is not a multiple of $N$.}
\item{Post-processing of the pseudo-probability distribution to get MAC.  Multiply all values in the resulting pseudo-probability distribution by $10^k$. Then retain the remainders of modulo $2^k$ to form a binary string as MAC. The bit length of MAC is $N\times k$.}
\end{enumerate}


CAQWBH is secure and efficient.  The CAQWBH-based MAC does not only reserve the advantages of the corresponding CAQWBH but also be resistant to the brute force search of a quantum computer, because the selectable space for $key_1$ is infinite and Aleph one. Therefore, the CAQWBH-based MAC executed on a quantum computer is resistant to attacks from a quantum computer.

Even it is executed on a classical computer, i.e. the selectable space for  $key_1$ and $key_2$  is finite, the CAQWBH-based MAC is so flexible that we can enlarge the keyspace to guarantee its security in current computing power.

\subsection{The CAQWBH-based Pseudo-random Number Generator}

Due to the uniform distribution and randomness of the hash value in the hash space,  the hash value of the CAQWBH satisfies the requirements of the pseudo-random number. Hence CAQWBH can be modified as a pseudo-random number generator.

 The generating process is described as follows:

\begin{enumerate}{}{}
\item{Select the parameters $(N, k, (\theta_1,\theta_2), \alpha_i)$. $\theta_1,\theta_2\in(0,\pi/2)$, $i\in \{0,\cdots, N-1\}$ and $\sum_{i}|\alpha_i|^2=1$.  $\theta_1,\theta_2$ are the parameters of the two coin operators respectively. Let the initial state be $|0\rangle(\sum_i\alpha_i|i\rangle)$.}
\item{Then run the CAQW  on $G_N$ one step to get the final state under the control of  a $N$-bit binary string, i.e. message. Then post-process the probability distribution to get a  pseudo-random number string. Multiply all values in the resulting probability distribution by $10^k$, then retain the remainders of modulo $2^k$ to form a binary string, which is a $Nk$-bit  pseudo-random number string.}
\item{If the pseudo-random number string is no longer enough, pick the last $N$ bits of the pseudo-random number string as the new message. Set the final state as the new initial state. Repeat step  4  and stick the  new pseudo-random number string to the end of the existing pseudo-random number string sequentially until the pseudo-random number string is long enough.}
\end{enumerate}

\section{Discussion}
\label{sec7}

CQW-based hash function is a kind of  novel hash function, which is based on the  quantum computation model quantum walk. They are safe, flexible, high-efficient. Furthermore, CQW-based hash function is compatible to be executed on a quantum computer or a classical computer.

In this paper, we focus on how to improve the  efficiency of the CQW-based hash function further. All existing CQW-based hash functions are controlled by one bit message in each step. To process message in batch amounts, CAQWBH is presented by using the timpe-position dependent controlled quantum walks on complete graphs with self-loops, which accelerate the hash processing dramatically. Besides, CAQWBH inherit all advantages of CQW-based hash function. The irreversible measurement and the modular arithmetic make sure that  CAQWBH is extremely safe. Statistical analysis prove the claim too. CAQWBH is flexible so that the hash value of different lengths is easy to attain by changing the parameters of the graph and the post-processing, i.e. $N$ and $k$. Furthermore, CAQWBH is a compatible hash function that it can be executed on a quantum computer or a classical computer. Taking the difficulty of the commercialization of quantum computers into consideration, CAQWBH is practical in current hardware technology.

In addition, extensions of CAQWBH, including CAQWBH-based message authentication code and CAQWBH-based pseudo-random number generator are introduced.

In summary, CAQWBH retains the superiorities of CQW-based hash function in security, flexibility,  and compatibility, while speeding up the efficiency.

\begin{acknowledgements}
This work is supported by NSFC (Grant Nos. 61901218, 61802025, 62071015), Natural Science Foundation of Jiangsu Province, China (Grant No. BK20190407), China Postdoctoral Science Foundation funded Project (Grant No. 2018M630557, 2018T110499), Jiangsu Planned Projects for Postdoctoral Research Funds (Grant No. 1701139B), National Key Research and Development Program of China (Grant No.2020YFB1005504).
\end{acknowledgements}

%
%



\end{document}